\documentclass[german,english]{bvm2020}

\def\articlenumber{2794}

\date{}
\title{Fully-automatic CT data preparation for interventional X-ray skin dose simulation}

\titlerunning{Fully-automatic CT data preparation}

\author{Philipp~Roser$^{1,3}$, Annette~Birkhold$^2$, Alexander~Preuhs$^1$, Bernhard~Stimpel$^1$, Christopher~Syben$^1$, Norbert~Strobel$^4$, Markus~Kowarschik$^2$, Rebecca~Fahrig$^2$, Andreas~Maier$^{1,3}$}

\authorrunning{Roser et al.}
\institute{%
 $^1$Pattern Recognition Lab, FAU Erlangen-N\"urnberg\\
 $^2$Siemens Healthcare GmbH, Forchheim \\
 $^3$Erlangen Graduate School in Advanced Optical Technologies (SAOT) \\
 $^4$Fakult\"at Elektrotechnik, HS f\"ur angewandte Wissenschaften W\"urzburg-Schweinfurt
 }
 
\email{philipp.roser@fau.de}

\begin{document}

\selectlanguage{english}

\maketitle

\begin{abstract}
Recently, deep learning (DL) found its way to interventional X-ray skin dose estimation.
While its performance was found to be acceptable, even more accurate results could be achieved if more data sets were available for training.  
One possibility is to turn to computed tomography (CT) data sets.
Typically, computed tomography (CT) scans can be mapped to tissue labels and mass densities to obtain training data.
However, care has to be taken to make sure that the different clinical settings are properly accounted for.
First, the interventional environment is characterized by wide variety of table setups that are significantly different from the typical patient tables used in conventional CT.
This cannot be ignored, since tables play a crucial role in sound skin dose estimation in an interventional setup, e.\,g., when the X-ray source is directly underneath a patient (posterior-anterior view).
Second, due to interpolation errors, most CT scans do not facilitate a clean segmentation of the skin border.
As a solution to these problems, we applied connected component labeling (CCL) and Canny edge detection to (a) robustly separate the patient from the table and (b) to identify the outermost skin layer.
Our results show that these extensions enable fully-automatic, generalized pre-processing of CT scans for further simulation of both skin dose and corresponding X-ray projections.
\end{abstract}

\section{Introduction}
Deep learning (DL) is a powerful technique for various applications, with medical X-ray imaging in general and computed tomography (CT) being no exception.
Whether for computer-aided diagnosis, semantic segmentation, or 3D reconstruction, DL elevated expected assessment metrics tremendously.
As Unberath \etal{} found, the data situation is excellent for many of these tasks -- with the exception of interventional imaging.
To exploit the abundance of existing high-quality and open access data libraries, they proposed a DL-fueled framework based on a poly-chromatic forward projector to generate realistic X-ray projections from CT scans\,\cite{Unberath:19:DeepDRR}.
While challenging tasks in the interventional environment, such as device tracking or learning-based trajectory planning, could be improved or, in some cases, even solved for the first time, skin dose monitoring and optimization does not benefit to the same extent.
Besides, DL-based 3D segmentation is often bound to a certain anatomic region and requires top tier hardware.

With increasing dose awareness in the interventional suite, physically-sound real-time tracking of skin dose during interventional procedures is desirable.
While being the gold standard, Monte Carlo (MC) simulation of X-ray photon transport typically suffers from high computational complexity.
Although hardware acceleration is widely applied for research purposes\,\cite{Badal:09:MCGPU}, it may not be available to the same degree in a clinical setting as the available hardware is usually used to support multiple imaging tasks.
Recently, DL was used to bypass the hardware bottleneck for dose estimation and promising first results could be reported for both high- and low-end hardware\,\cite{Roser:19:DoseLearning}.
However, the generation of ground truth data is a tedious task due to (a) the high computational complexity of general purpose MC codes and (b) the lack of open-access digital phantoms.
To increase the number of data sets for the generation of ground-truth data using MC simulations, CT scans can be mapped to corresponding tissue types and densities using thresholds.
Unfortunately tables, mattresses, and blankets are acquired well in most CT protocols, whereas the skin can not be segmented distinctly.
To account for the many different setups (tables, patient preparation) in an interventional environment, a data preparation pipeline capable of removing tables while finding an adequate skin segmentation is needed.
To this end, we propose a framework that extends the conventional approach of mapping Hounsfield units (HU) to tissue labels\,\cite{Schneider:00:HU2Tissue} and densities by unsupervised image processing techniques. 
This ensures smooth tissue distributions, homogeneous skin segmentation, and removes non-patient objects such as tables, blankets, and mattresses. 
Our framework will be made open source after publication.

\section{Methods}
\begin{figure}
    \centering
    \includegraphics[width=\linewidth]{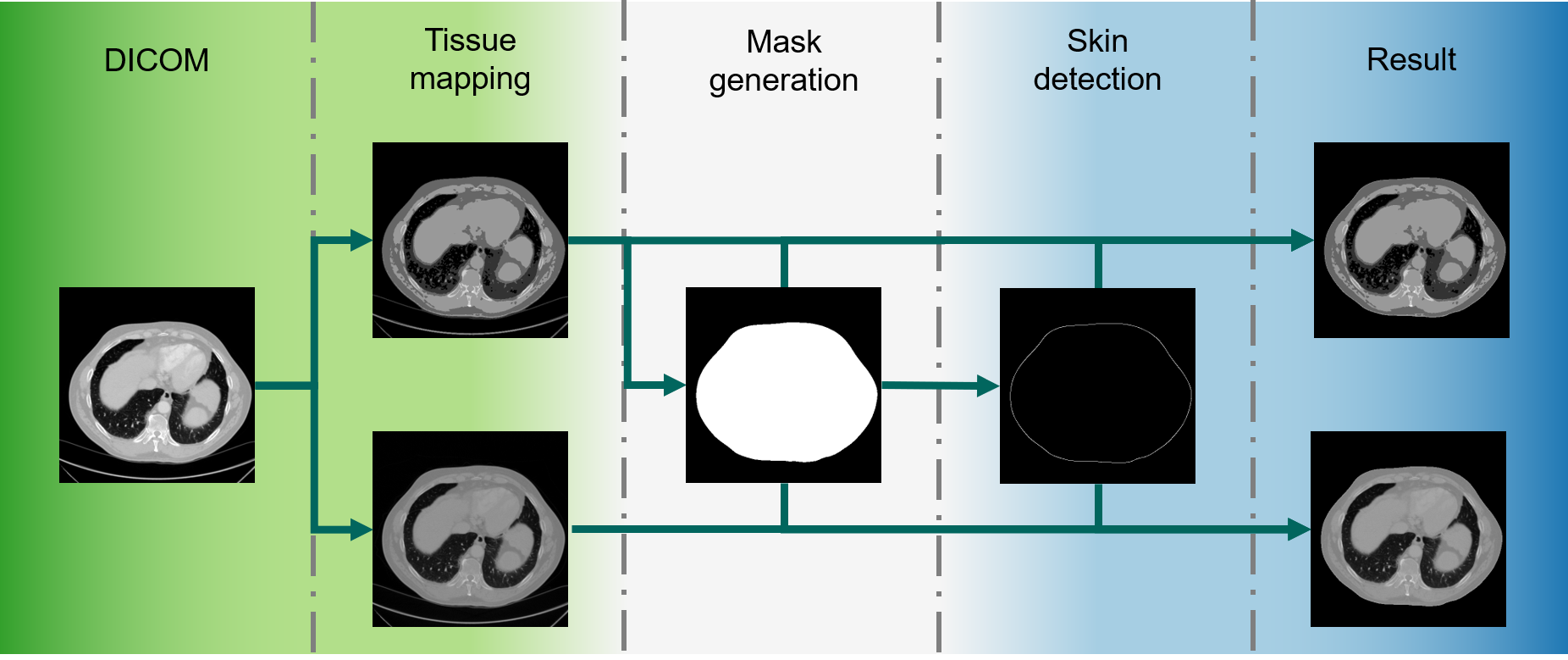}
    \caption{Flow diagram of the proposed method. Based on the HU values of a CT scan, tissue labels and associated densities are derived. Then a mask of the body is generated using connected component analysis. Finally, the skin is labeled using edge detection on the mask.}
    \label{fig:pipeline}
\end{figure}
The outline of the proposed data preparation pipeline is depicted in Fig.\,\ref{fig:pipeline}.
The integer pixel values $\vec{I}_\text{r} \in \mathbb{Z}^3$ are mapped to HU units using the specified rescale intercept $a_\text{r} \in \mathbb{Z}$ and slope $b_\text{r} \in \mathbb{R}$ according to $\vec{I}_\text{HU} = \left[a_\text{r} + b_\text{r} \vec{I}_\text{r}\right] \in \mathbb{Z}^3$.
Here `$[\cdot]$' rounds to the next integer number.

\subsection{Tissue mapping}
\begin{figure}
    \begin{tabular}{@{\extracolsep{10pt}} cc @{}}
        \includegraphics[width=0.45\linewidth]{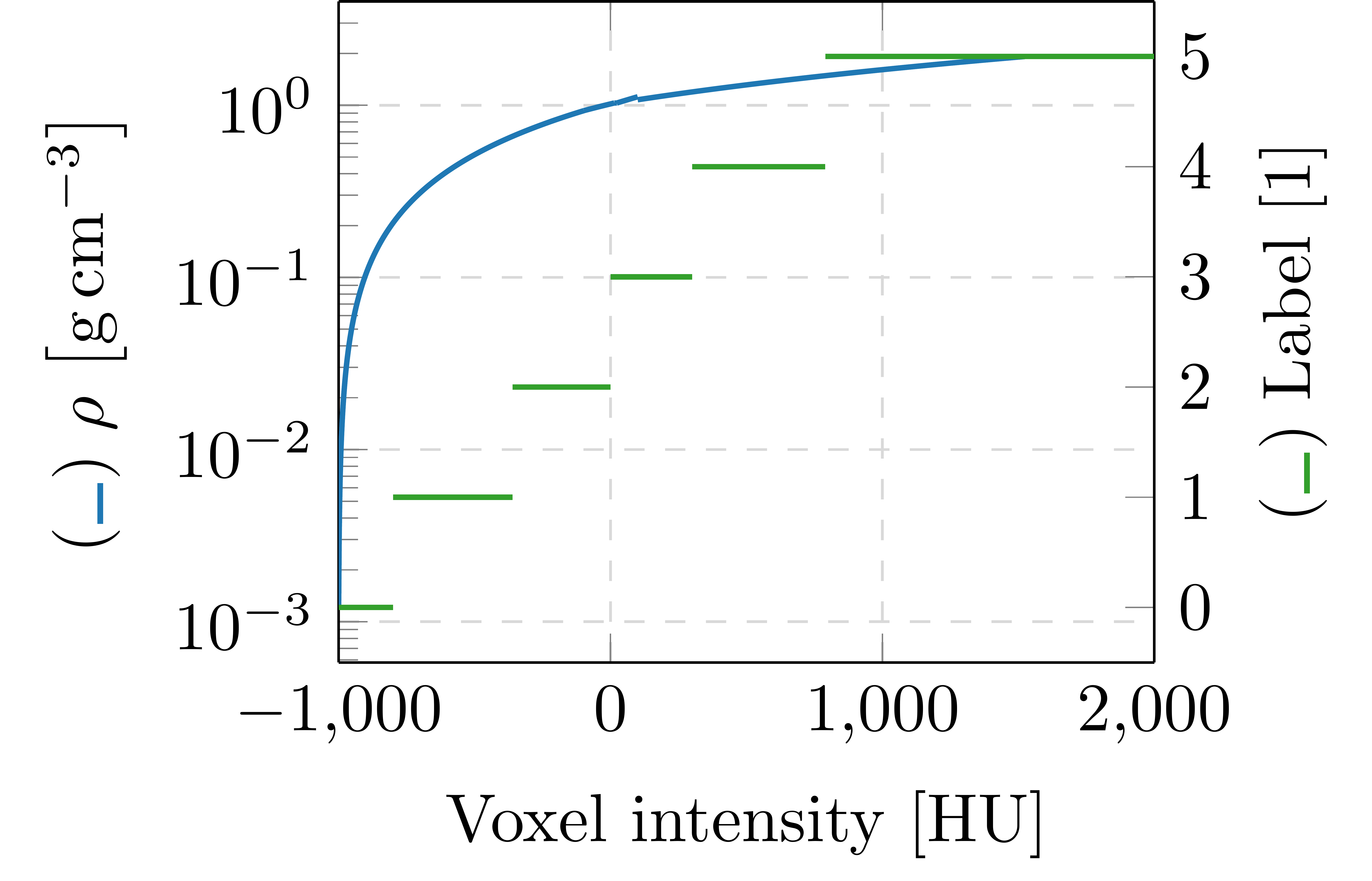} & 
        \fbox{\includegraphics[width=0.45\linewidth,]{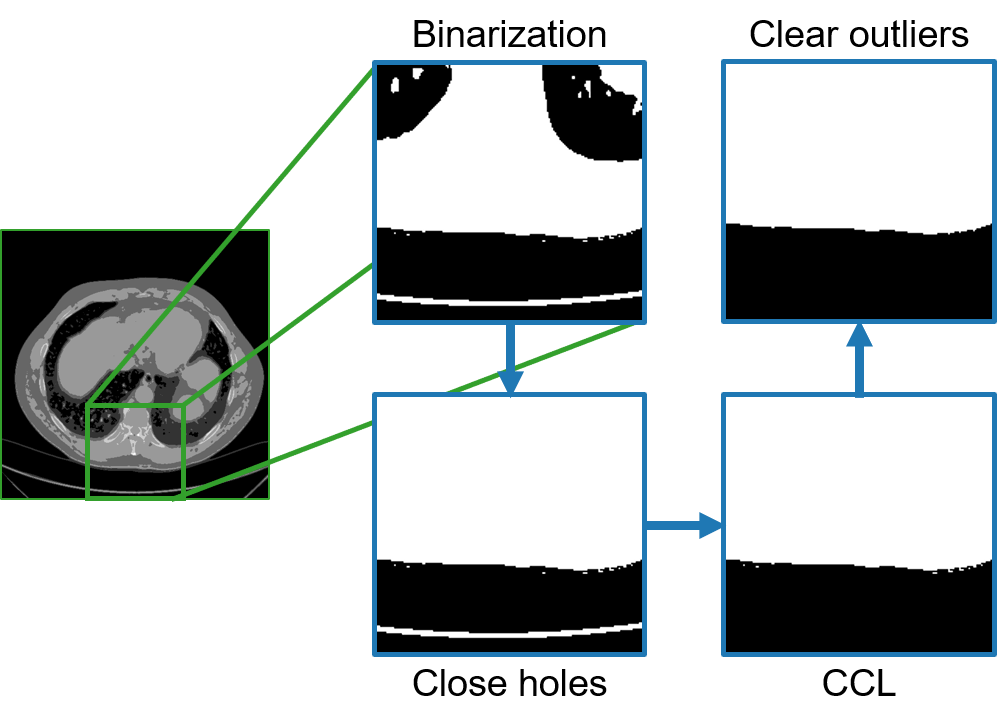}} \\
        (a) Tissue mapping & (b) Mask generation \\
    \end{tabular}
    \caption{HU values are mapped to mass densities $\rho$ and tissue labels using piece-wise linear and constant transfer functions (a). From the initial tissue labels, a mask is generated. 
    To this end, we apply binarization and close holes found in the lung or esophagus.
    Afterwards, we separate objects using connected component labeling (CCL).
    Finally leftover outliers are cleared (b).}
    \label{fig:mapmask}
\end{figure}
To preserve inhomogeneities in the human body that might play a crucial role in realistic data generation, $\vec{I}_\text{HU}$ is mapped to mass densities $\vec{\rho} \in \mathbb{R}^3$ using piece-wise linear transforms $\rho(I_\text{HU}) = a_\rho + b_\rho I_\text{HU}$\,, with $a_\rho, b_\rho \in \mathbb{R}$\,\cite{Schneider:00:HU2Tissue}.
To obtain tissue labels $\vec{L} \in \mathbb{N}^3$, associated HU values are mapped based on a piece-wise constant function.
Both transforms are plotted in Fig.\,\ref{fig:mapmask}\,a.
Note that due to overlapping HU value ranges for certain structures, such as cancellous bone and other organs, the tissue -- and to some extent the density -- mapping is not bijective in general.
However, in these value ranges, the different tissue types behave similarly.
Therefore the tissue label transfer function is defined to provide a meaningful trade-off between all overlapping HU value ranges.
In total, we differentiate between lung tissue (1), adipose tissue (2), soft tissue (3), cancellous bone (4), cortical bone (5), and air (0).
For convenience, the labels are ordered based on their corresponding nominal mass densities.
In addition, outlier pixels in large homogeneoues regions are removed using a median-based hot pixel detection algorithm.

\subsection{Mask generation}
Our approach to generate the patient mask $\vec{M} \in \{0,1\}^3$ is illustrated in Fig.\,\ref{fig:mapmask}\,b.
Initially, the tissue labels are binarized, where each non-zero label is assigned to one.
To improve the performance of subsequent steps, zero-areas within closed contours, such as the lung or esophagus, are closed.
Then we apply connected component labeling (CCL)\,\cite{He:17:CCL} to separate, e.\,g., the table from the patient body.
The patient is identified by the largest connected cluster, thus all other clusters are assigned to zero.
Eventually, the same outlier detection as previously is applied to remove possible leftover hot pixels in the air or near to the patient surface.

\subsection{Skin detection}
Due to artifacts, interpolation, and tissue ambiguities, the skin is rarely identified as a homogeneous soft tissue region after the initial processing steps.
Instead, we rather find a mixture of lung, adipose, and soft tissue.
While this is negligible for realistic X-ray image generation, a unique skin surface label is desirable for skin entrance and back-scatter dose estimation.
To identify the skin surface, we apply the Canny edge detector\,\cite{Canny:86:Edge} with standard deviation $\sigma=1$ to the binary patient mask yielding a one-voxel thick edge around the patient surface $\vec{S} \in \{0,1\}^3$.
\begin{figure}
    \centering
    \includegraphics[width=\textwidth]{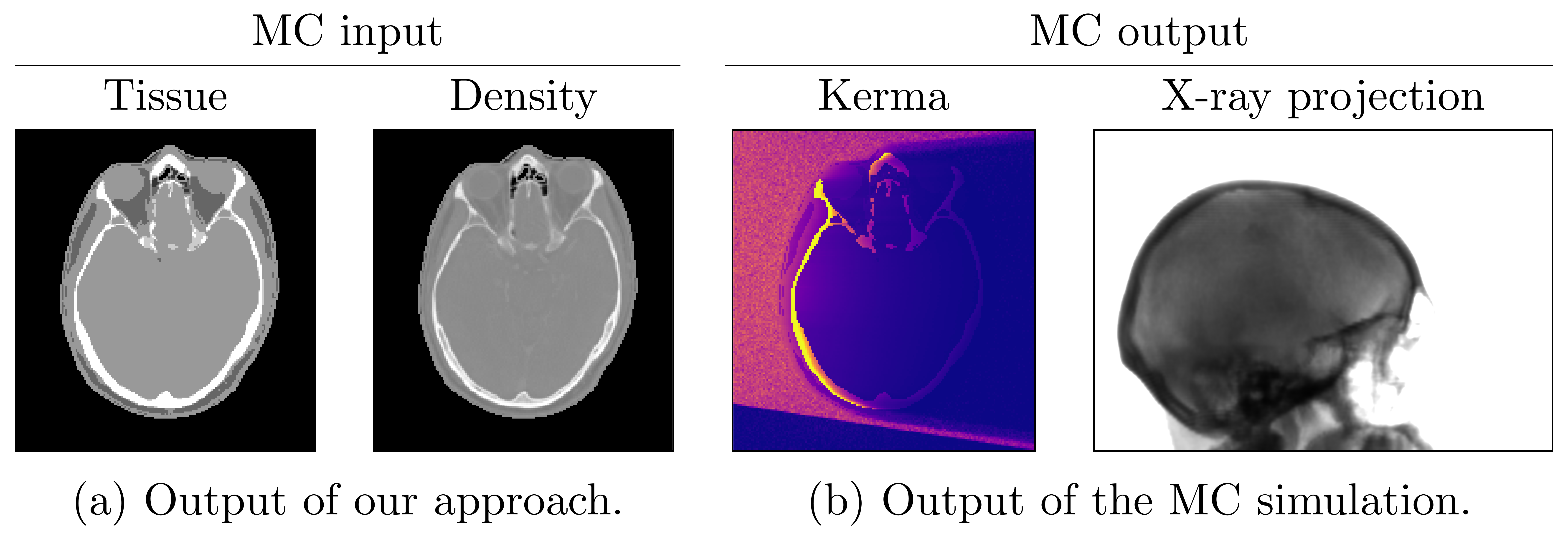}
    \caption{Sample inputs and outputs of the MC-GPU code of a CT scan pre-processed by the proposed data pre-processing framework.}
    \label{fig:example}
\end{figure}
\begin{figure}
    \centering
    \begin{tabular}{@{\extracolsep{10pt}} cccc @{}}
         & Average DICOM & Average label & Average density \\
        \rotatebox{90}{HNSCC-3DCT-RT} &
        \includegraphics[width=0.25\linewidth]{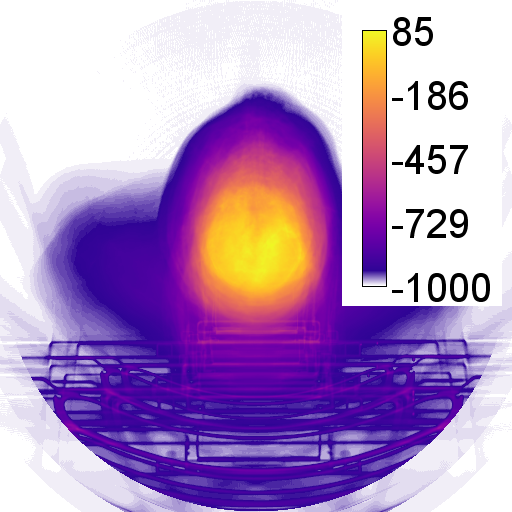} &
        \includegraphics[width=0.25\linewidth]{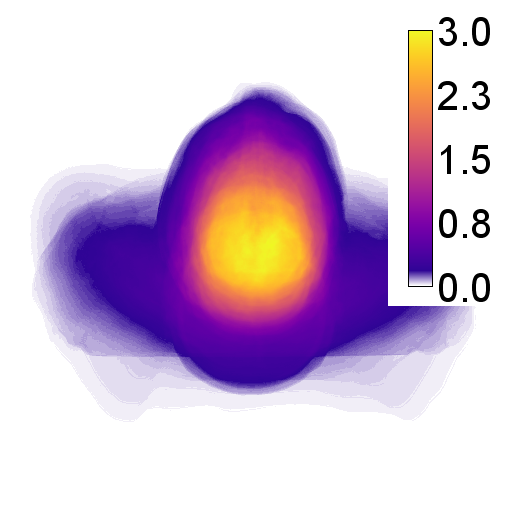} &
        \includegraphics[width=0.25\linewidth]{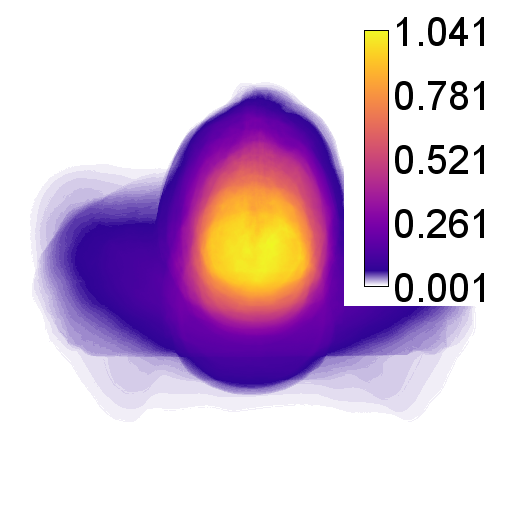} \\
        \rotatebox{90}{CT Lymph Nodes} &
        \includegraphics[width=0.25\linewidth]{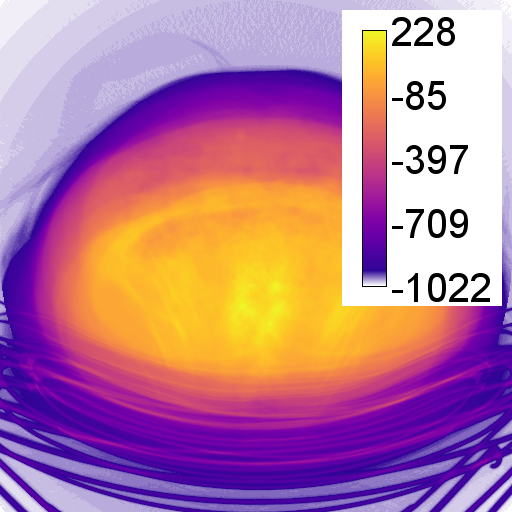} &
        \includegraphics[width=0.25\linewidth]{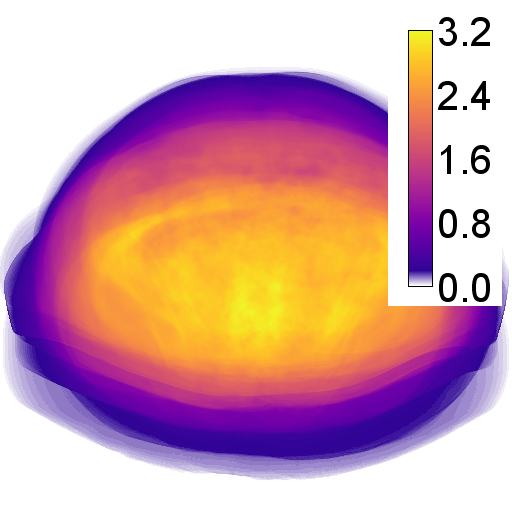} &
        \includegraphics[width=0.25\linewidth]{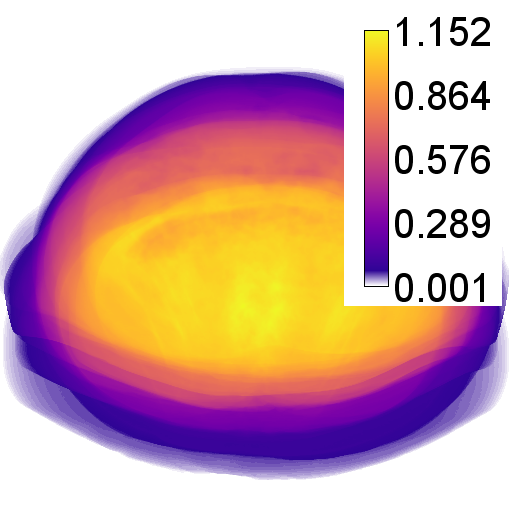} \\
    \end{tabular}
    \caption{Average input DICOM, generated labels, densities of 10 volumes from two publicly available datasets. By visual inspection it is evident that table and artifacts get removed completely while the overall correlation is preserved.}
    \label{fig:averages}
\end{figure}

\subsection{Composition}
Based on the mask, $\vec{M}$, the skin surface, $\vec{S}$, the tissue labels, $\vec{L}$,  and the densities, $\vec{\rho}$, the final labeled data set, $\vec{L}^\prime$, and density distribution, $\vec{\rho}^\prime$, are computed as 
\begin{align}
    \vec{L}^\prime & = (\vec{M} - \vec{S}) \odot \vec{L} + L_\text{soft} \vec{S}\,, \\
    \vec{\rho}^\prime & = (\vec{M} - \vec{S}) \odot \vec{\rho} + \rho_\text{soft} \vec{S}\,,
\end{align}
where $L_\text{soft}$ and $\rho_\text{soft}$ denote the label and nominal mass density for soft tissue, respectively.
The operator `$\odot$' denotes the element-wise multiplication.
For convenience, the pipeline also provides an interface to the {penEasy 2008} voxel volume format that is used in the popular Penelope\,\cite{Sempau:97:Penelope} and MC-GPU\,\cite{Badal:09:MCGPU} codes.

\section{Results}
Quantitatively assessing the performance of the proposed data preparation method is inherently difficult as there is currently no ground truth data available.
To still obtain some insights into the merits of our work at this point in time, we study how well the anatomical shape is preserved and whether table gets removed without impairing the neighboring patient outline.
Figure\,\ref{fig:example}\,a shows sample outputs of our proposed approach.
We found that the tissue labels obtained were homogeneously distributed.
In addition, the estimated tissue densities facilitated the generation of realistic X-ray projection images.
This observation is further substantiated by the perceptually realistic kerma distribution and X-ray projection estimated using MC simulation, which are shown in Fig.\,\ref{fig:example}\,b.
To illustrate the performance of the table removal, Fig.\,\ref{fig:averages} shows average intensity projections of 10 CT scans along the longitudinal axis taken from the {HNSCC-3DCT-RT} head\,\cite{Bejanero:18:Heads} and the {CT Lymph Nodes} torso\,\cite{Roth:15:Lymphs} datasets provided in {The Cancer Imaging Archive} (TCIA)\,\cite{Clark:13:TCIA}.
Evidently, the average shapes are well preserved, while the table and acquisition-induced artifacts were removed completely.

\section{Discussion}
We presented a fully-automatic, robust, generalized CT scan processing pipeline.
It can be used to generate input for MC simulation codes without any need for manual interaction and thus making it attractive for large scale data generation.
Thanks to the proposed method, new CT-based digital models can be used to extend our existing training set.
Especially the removal of tables as used for conventional spiral CT scans and the inclusion of a distinct skin voxel layer makes it possible to transfer the data to interventional setups and use it there for skin dose estimation.
We also found that our method produces anatomiccally consistent results for both head and torso scans without adapting any parameters.

Still, the pipeline certainly has limitations.
First, overlapping HU value ranges of different tissue types -- most prominently soft tissue, brain tissue, and cancellous bone -- needs to be addressed in future iterations.
This could be achieved using (a) a locally restricted CCL or (b) a DL approach.
Second, the skin modelling could be enhanced by differentiating between sub-layers of skin.
To this end, a pre-defined skin template could be matched with respect to the normal vector of the patient surface.
Third, especially in head CTs, metal artifacts occur and can currently not be dealt with sufficiently well.
One way to solve this issue could be an inpainting approach based on template matching or generative adversarial networks.

~\\
{\bf Disclaimer:} The concepts and information presented in this paper are based on research and are not commercially available.

\bibliographystyle{bvm2020}

\bibliography{2794}
\marginpar{\color{white}E\articlenumber} 
\end{document}